\documentstyle[12pt]{article}
\def\Re{{\cal R \mskip-4mu \lower.1ex \hbox{\it e}\,}}
\def\Im{{\cal I \mskip-5mu \lower.1ex \hbox{\it m}\,}}

\def\etal{{\it et al.}}

\def\sub#1{_{\lower.25ex\hbox{$\scriptstyle#1$}}}
\def\sul#1{_{\kern-.1em#1}}
\def\sll#1{_{\kern-.2em#1}}
\def\sbl#1{_{\kern-.1em\lower.25ex\hbox{$\scriptstyle#1$}}}
\def\ssb#1{_{\lower.25ex\hbox{$\scriptscriptstyle#1$}}}
\def\sbb#1{_{\lower.4ex\hbox{$\scriptstyle#1$}}}

\def\tev{\,{\rm TeV}}
\def\gev{\,{\rm GeV}}

\def\to{\rightarrow}
\def\mh{\ifmmode m_{h^0} \else $m_{h^0}$\fi}
\def\mch{\ifmmode m_{H^\pm} \else $m_{H^\pm}$\fi}
\def\mt{\ifmmode m_t\else $m_t$\fi}
\def\mc{\ifmmode m_c\else $m_c$\fi}
\def\mz{\ifmmode M_Z\else $M_Z$\fi}
\def\mw{\ifmmode M_W\else $M_W$\fi}
\def\mws{\ifmmode M_W^2 \else $M_W^2$\fi}
\def\mhs{\ifmmode m_H^2 \else $m_H^2$\fi}
\def\mzs{\ifmmode M_Z^2 \else $M_Z^2$\fi}
\def\mts{\ifmmode m_t^2 \else $m_t^2$\fi}
\def\mcs{\ifmmode m_c^2 \else $m_c^2$\fi}
\def\mchs{\ifmmode m_{H^\pm}^2 \else $m_{H^\pm}^2$\fi}
\def\ztwo{\ifmmode Z_2\else $Z_2$\fi}
\def\zone{\ifmmode Z_1\else $Z_1$\fi}
\def\mtwo{\ifmmode M_2\else $M_2$\fi}
\def\mone{\ifmmode M_1\else $M_1$\fi}
\def\tb{\ifmmode \tan\beta \else $\tan\beta$\fi}
\def\xw{\ifmmode x\sub w\else $x\sub w$\fi}
\def\ch{\ifmmode H^\pm \else $H^\pm$\fi}
\def\lum{\ifmmode {\cal L}\else ${\cal L}$\fi}
\def\inpb{\ifmmode {\rm pb}^{-1}\else ${\rm pb}^{-1}$\fi}
\def\infb{\ifmmode {\rm fb}^{-1}\else ${\rm fb}^{-1}$\fi}
\def\epem{\ifmmode e^+e^-\else $e^+e^-$\fi}
\def\ppb{\ifmmode \bar pp\else $\bar pp$\fi}

\newskip\zatskip \zatskip=0pt plus0pt minus0pt
\def\matth{\mathsurround=0pt}
\def\lsim{\mathrel{\mathpalette\atversim<}}
\def\gsim{\mathrel{\mathpalette\atversim>}}
\def\atversim#1#2{\lower0.7ex\vbox{\baselineskip\zatskip\lineskip\zatskip
  \lineskiplimit 0pt\ialign{$\matth#1\hfil##\hfil$\crcr#2\crcr\sim\crcr}}}

\renewcommand{\thefootnote}{\fnsymbol{footnote}}

\begin{document} \begin{titlepage}

\rightline{\vbox{\halign{&#\hfil\cr
&ANL-HEP-PR-92-110\cr
&OTIS-501\cr
&November 1992\cr}}}
\vspace{0.5in}
\begin{center}

{\Large\bf
Can $b\to s\gamma$ Close the Supersymmetric Higgs Production Window?}
\medskip
\vskip0.5in

\normalsize JOANNE L.\ HEWETT
\\ \smallskip
\medskip

High Energy Physics Division, Argonne National Laboratory,
Argonne, IL 60439\\
\smallskip
and\\
\smallskip
Institute of Theoretical Science, University of Oregon, Eugene, OR 97403\\
\smallskip

\end{center}
\vskip1.0in

\begin{abstract}

We show that the present limit from CLEO on the inclusive decay
$b\to s\gamma$ provides strong constraints on
the parameters of the charged Higgs
sector in two-Higgs-Doublet-Models.  Only a slight improvement in the
experimental bound will exclude the region in the Supersymmetric Higgs
parameter space which is inaccessible to collider searches.

\end{abstract}

\renewcommand{\thefootnote}{\arabic{footnote}} \end{titlepage}


In the Standard Model (SM) of electroweak interactions, spontaneous symmetry
breaking is achieved through a single Higgs doublet, which then generates
masses for both the gauge boson and fermion sectors.  However,
many extensions of the SM predict the existence of
two Higgs doublets\cite{hhg}.  The physical spectrum of these models consists
of three neutral Higgs scalars, two of which are CP-even ($h^0, H^0$) while one
is CP-odd ($A^0$), and two charged Higgs scalars (\ch).  In general,
tree-level flavor changing neutral currents are present in
two-Higgs-Doublet Models (2HDM), but can naturally be
avoided if each fermion type receives mass from the vacuum expectation value
(vev) of a single doublet.
In one such model, hereafter referred to as Model\ I, one doublet ($\phi_2$)
provides masses for all fermions and the
other doublet ($\phi_1$) decouples from the fermion sector.
This model predicts distinctive phenomenology and has recently been
revived in the literature\cite{modeli}.
In a second model (Model\ II), $\phi_2$ gives mass to the up-type quarks, while
the down-type quarks and charged leptons receive their masses from $\phi_1$.
Supersymmetry and many axion theories predict couplings of the type present
in Model\ II.

Many different types of experiments may reveal the existence of
physics beyond the SM.
In the case of 2HDM, one could hunt for the enlarged Higgs spectrum
directly in both \epem\ and hadron colliders.  The strongest direct search
limits on these particles are provided by LEP\cite{lep}; in particular,
the mass of the \ch\ is restricted to be $\gsim \mz/2$ in all 2HDM.
It is also possible that such new physics may manifest itself indirectly in
low-energy phenomena.
Here, we will show that powerful bounds on the charged Higgs
sector in 2HDM arise from the radiative b-quark decay, $b\to s\gamma$, and that
these constraints surpass those from direct collider searches.  As we will see
below, the application of these bounds to the particular case of supersymmetric
models, could close the window of parameter space where there are no
detectable Higgs signatures at colliders.  A $90\%$
C.L. upper bound on the branching fraction for this mode, $B(b\to s\gamma)
< 8.4\times 10^{-4}$,  has been obtained by the CLEO Collaboration\cite{cleo}
via an examination of the inclusive photon spectrum in
$B$-meson decays.  Recent detector refinements coupled with increasing
integrated luminosity leads us to anticipate that either the current limit will
be strengthened, or the decay may actually be observed in the near future.

Before presenting our results, we first briefly describe the nomenclature of
2HDM.  Each doublet obtains a vev $v_i$, subject only to the constraint that
$v_1^2 +v^2_2=v^2$, where $v$ is the usual vev present in the SM.
The charged Higgs interactions with fermions are then governed by the
Lagrangian
\begin{eqnarray}
{\cal L} & = & {g\over 2\sqrt 2 \, M_W} H^\pm [ V_{ij}^{\phantom9}
m_{u_i} A_u \bar u_i (1-\gamma_5) d_j  \\
         &   & \mbox{} + V_{ij} m_{d_j} A_d \bar u_i (1+\gamma_5) d_j +
m_\ell A_\ell \bar\nu(1 + \gamma_5)\ell ] + {\rm h.c.} \,, \nonumber
\end{eqnarray}
where $g$ is the usual SU(2)$_L$ coupling constant and $V_{ij}$ is the
appropriate element of the Kobayashi Maskawa (KM) matrix.  In Model~I,
$A_u=\cot\beta$ and $A_d=A_\ell=-\cot\beta$, and in Model~II,
$A_u=\cot\beta$ and $A_d=A_\ell=\tan\beta$, where $\tan\beta\equiv v_2/v_1$
is the ratio of vevs.  In a general 2HDM, the mass $m_{H^\pm}$ and \tb\ are
{\it a priori} free parameters, as are the masses of all the neutral Higgs
fields.  However, in supersymmetric models, mass relationships exist between
the various Higgs scalars.  At tree-level,in such models,
only two parameters are required
to fix the masses and couplings of the entire scalar sector, but once radiative
corrections are included\cite{radcorr}, the values of the top-quark and squark
masses also need to be specified.

The transition $b\to s\gamma$ proceeds  through electromagnetic penguin
diagrams, which involve the top-quark, together with a \ch\ or SM $W^\pm$
boson in the loop.  The expression for the branching fraction in 2HDM is given
in the literature\cite{bsg}.
To fully appreciate our results, it is important to observe the \tb\
dependence in the transition amplitude for both models.
At the $W$ scale the coefficients of the operators which mediate this
transition
take the generic form
\begin{equation}
c_i(\mw)=A_W(\mt^2/\mw^2)+\lambda A^1_H(\mt^2/\mch^2)+
{1\over\tb^2}A^2_H(\mt^2/\mch^2)  \,,
\end{equation}
where $\lambda = -1/\tb^2,\ +1$\ in Models I and II, respectively.  $A_W$
corresponds to the SM amplitude and $A_H^{1,2}$ represent the \ch\
contributions; their analytic form is given for each contributing operator in
Ref.\ 6.  We employ the explicit form of the QCD corrections  as stated in
Eqns.\ (4.2) and (4.6) of Grinstein \etal\cite{bsg}.
In order to evaluate these corrections, we use the 3-loop expression for
$\alpha_s$ and fit the value of the QCD scale $\Lambda$ to obtain consistency
with measurements\cite{lep} of $\alpha_s(\mz^2)$ at LEP.  In the SM,
this procedure yields $B(b\to s\gamma)=(2.56 - 3.94) \times 10^{-4}$ for
\mt\ in the range $90 - 200\gev$.

The dependency of the branching fraction
on \mt, \mch, and \tb\ in both Models\ I and II is explicitly presented
in Ref.\ 7.  Enhancements over the SM rate only occur in Model\ I for small
values of \tb.  This is due to the fact that the \ch\ contributions to
the amplitude are always scaled as $\cot^2\beta$, as seen in Eqn.\ 2.  We
also note that a destructive interference between the \ch\ and $W$
contributions
is possible for some values of the parameters due to the minus sign in Eqn.\ 2.
In Model\ II, large enhancements also appear for small values of \tb, but
more importantly, the branching fraction is found to {\it always} be larger
than that of the SM.  This occurs independently of the value of \tb\ due to the
presence of the term $A^1_H$.
For certain ranges of the model parameters, the resulting value
of $B(b\to s\gamma)$ exceeds the CLEO bound, and consistency with this limit
thus excludes part of the $\mch - \tb$ plane for a fixed value of \mt.  This
is shown in Fig.\ 1 for both models, where the excluded region lies to the left
and beneath the curves.  We see that in Model\ I only a small region of the
parameter space is excluded, regardless of the value of \mt, however, in
Model\ II a lower limit on \mch\ is obtained for all values of \tb\ for the
reasons discussed above.  For example, with $\mt=150\gev$, we find that
$\mch>110\gev$ at large \tb, and that even
stronger bounds on \mch\ result for values of $\tb\lsim 1$.  If the CLEO
limit were to substantially improve, the constraints in Model\ II would
strengthen drastically, while those in Model\ I would be essentially unchanged.
In performing our calculations, we have tried to be as conservative as
possible.
We scale $\Gamma(b\to s\gamma)$ to the $b$-quark
semi-leptonic decay width employing
both QCD and phase space corrections (taking $m_c/m_b=0.3$) to $b\to
ce\bar\nu_e$, and take $B(b\to ce\bar\nu_e)=10.8\%$ (Ref.\ 8).  This
procedure removes the sensitive dependence of the $b\to s\gamma$ rate to the
values of the KM matrix elements.  The
precise value of $m_b$ (we set $m_b=5\gev$) that is used in the QCD corrections
and phase space factors can have a sizable effect.  For example,
if $m_b=4.5\gev$ were taken, the bound on \mch\ in Model\ II would increase
to $165\gev$ for $m_t=150\gev$ in the large \tb\ limit.  QCD corrections
which include the effects of separate mass scales for $m_t>\mw$ are
found\cite{cho} to increase the radiative decay rate by a few
percent, but have not
been used here, since a separate scale for \mch\ should also be taken into
account.  However, we expect that this effect would not be too large
since the running of $\alpha_s$ is slow between pairs of heavier mass scales.

To show the power of this indirect search technique, we present in Fig.\ 2
the branching fraction for Model\ II in the limit of large \tb\ as a function
of \mt\ with $\mch=\mt-m_b$.  If the actual value (or future upper bound) for
the branching fraction were to lie below the solid curve, then the decay
$t\to b\ch$ is kinematically forbidden for a particular value of \mt.
For example, the present CLEO result implies that $t\to b\ch$ is already
excluded for $\mt\gsim 215\gev$.  We see that if the CLEO upper limit were to
improve to $6\times 10^{-4}$, then this decay mode of the top-quark would
be excluded for all values of the model parameters.  Here, we made use of the
large \tb\ limit, since it minimizes the \ch\ contributions to $b\to s\gamma$.

In the supersymmetric case, the bounds shown in Fig.\ 1b are more
conventionally displayed as an allowed region in the $\tb - m_A$\ plane, where
$m_A$ is the mass of the CP-odd field.  This is displayed in
Fig.\ 3a for various values of \mt, where the radiative corrections to the
SUSY mass relations have been employed assuming $M_{SUSY}=1\tev$.
We find that our results depend only weakly on
the exact value chosen for the squark masses.
For $\mt=150\gev$, the excluded region is comparable to what can be
explored\cite{collider} by LEP I and II.  The region of supersymmetric
parameter space which remains is exactly that in which the lightest CP-even
scalar behaves like the SM Higgs.  Figure 3b shows the growth in
the size of the excluded region if the CLEO bound were to improve to
$B(b\to s\gamma) < 4,5,6,~{\rm or}~7\times 10^{-4}$, assuming $\mt=150\gev$.
If $B<6\times 10^{-4}$ then the window of parameter space left
uncovered\cite{collider} by both \epem\ and hadron collider searches would be
{\it excluded}.  This would imply that the SSC/LHC could cover the entire
remaining allowed supersymmetric Higgs parameter region.

We note that in supersymmetric theories, other super-particles can also
contribute to the one-loop decay $b\to s\gamma$, and generally lead to a
further enhancement in the rate\cite{bert}.  However, a complete examination
of the full supersymmetric parameter space needs to be performed to determine
if
supersymmetric contributions {\it always} enhance
the branching fraction.  If that were
the case, then the limits presented in this paper could be strengthened in
the supersymmetric version of the 2HDM.  However, if some range of
parameter values yielded a destructive interference in the $b\to s\gamma$
amplitude, the bounds in the supersymmetric 2HDM could become weaker.

In conclusion, we have shown that the decay $b\to s\gamma$ is by far the most
restrictive process in constraining the parameters of the charged Higgs
sector in 2HDM, yielding bounds which are stronger than those from other
low-energy processes and from direct collider searches.
We anxiously await future results from the CLEO experiment.

\vskip.25in
\centerline{ACKNOWLEDGEMENTS}

The author thanks N.G. Deshpande and T.G. Rizzo for fruitful discussions and
the Institute of Theoretical Science at the
University of Oregon for
their hospitality and helpfulness during the course of this work.
This work was supported in
part by the U.S.~Department of Energy
under contract W-31-109-ENG-38.

\newpage

%
\def\MPL #1 #2 #3 {Mod.~Phys.~Lett.~{\bf#1},\ #2 (#3)}
\def\NPB #1 #2 #3 {Nucl.~Phys.~{\bf#1},\ #2 (#3)}
\def\PLB #1 #2 #3 {Phys.~Lett.~{\bf#1},\ #2 (#3)}
\def\PR #1 #2 #3 {Phys.~Rep.~{\bf#1},\ #2 (#3)}
\def\PRD #1 #2 #3 {Phys.~Rev.~{\bf#1},\ #2 (#3)}
\def\PRL #1 #2 #3 {Phys.~Rev.~Lett.~{\bf#1},\ #2 (#3)}
\def\RMP #1 #2 #3 {Rev.~Mod.~Phys.~{\bf#1},\ #2 (#3)}
\def\ZP #1 #2 #3 {Z.~Phys.~{\bf#1},\ #2 (#3)}

\newpage

%
\noindent{\bf Figure Captions}
\begin{itemize}

\item[Figure 1.]{The excluded regions in the $\mch - \tb$ plane for various
values of \mt, resulting from the present CLEO bound in (a) Model\ I and
(b) Model\ II.  In each case, from top to bottom, the solid
(dashed dot, solid, dotted, and dashed)  curve corresponds
to $m_t=210 (180, 150, 120,\ {\rm and}\ 90)\gev$.
The excluded region lies to the left and below each curve.}
\item[Figure 2.]{$B(b\to s\gamma)$ as a function of \mt, with $\mch=\mt-
m_b$ in the large \tb\ limit in Model\ II.}
\item[Figure 3.]{(a)  The excluded region from the present CLEO limit in the
$\tb - m_A$ plane for various values of \mt\ as indicated.  (b)  Excluded
regions for $\mt=150\gev$, if the CLEO bound was improved to $B<
4,5,6~{\rm or}~7\times 10^{-4}$. In each case, the excluded region lies to
the left and below each curve.}
\end{itemize}

\end{document}